# Optical spectra and structure of $CdP_4$ nanoclusters fabricated by incorporation into zeolite and laser ablation

O.A. Yeshchenko, I.M. Dmitruk, S.V. Koryakov, M.P. Galak

Physics Department, National Taras Shevchenko Kyiv University,

2 Akademik Glushkov prosp., 03022 Kyiv, Ukraine

yes@univ.kiev.ua

Corresponding author: O.A.Yeshchenko.

Tel.: +380-44-5264587; Fax: +380-44-5264036; E-mail: yes@univ.kiev.ua

$CdP_4$ nanoclusters were fabricated by incorporation into the pores of zeolite Na-X and by deposition of the clusters onto a quartz substrate using laser ablation-evaporation technique. Absorption and photoluminescence (PL) spectra of $CdP_4$ nanoclusters in zeolite were measured at the temperatures of 4.2, 77 and 293 K. Both absorption and PL spectra consist of two blue shifted bands. We performed DFT calculations to determine the most stable clusters configuration in the size region up to size of the zeolite Na-X supercage. The bands observed in absorption and PL spectra were attributed to emission of $(CdP_4)_3$ and $(CdP_4)_4$ clusters with binding energies of 3.78 eV and 4.37 eV per atom respectively. The Raman spectrum of $CdP_4$ clusters in zeolite proved the fact of creation of $(CdP_4)_3$ and $(CdP_4)_4$ clusters in zeolite pores. The PL spectrum of $CdP_4$ clusters produced by laser ablation consists of single band that was attributed to emission of $(CdP_4)_4$ cluster.

1. Introduction

There are several methods for semiconductor nanoparticles fabrication, e.g. fabrication of nanoparticles in solutions,[1] glasses[2] or polymers.[3] However, it is not easy to control the size distribution of small nanoparticles with countable number of atoms (so called clusters) using these methods. Matrix method based on the incorporation of materials into the 3D regular system of voids and channels of zeolite crystals could be an alternative technique for fabrication of semiconductor nanoclusters with controllable size distribution.[4,5] Many works have been reported on the nanoclusters incorporated into zeolite pores: semiconductors[6-9], metals[10,11] and polymers.[12,13] The subnanometer and nanometer clusters are very interesting as they are intermediate between the molecules and the typical nanocrystals. Usually, the structure of nanoclusters is different from the structure of nanocrystals, which resembles the structure of bulk crystals.[14] As a rule, the calculation methods of the electronic states structure of nanocrystals that are based on the effective mass approximation are not applicable for clusters. Thus, nanoclusters are very interesting objects as their structure, electronic and vibrational properties are quite different from the crystalline nanoparticles. Zeolites provide the opportunity to obtain extremely small clusters in the pores with diameter up to 15 Å. Zeolites are crystalline alumosilicates with cavities which diameter can vary in the range from 7 to15 Å. The size of the cage depends on the kind of alumosilicate framework, ratio Si/Al, origin of ion-exchanged cations, which stabilise negative charge of framework, etc. Zeolite Na-X, which has been used in the present work has Si/Al ratio equal 1, Fd3m symmetry and two types of cages: one is sodalite cage – truncated octahedron with diameter 8 Å and supercage, which is formed by the connection of sodalites in diamond-like structure with the diameter of about 13 Å.[15] All cages are interconnected by shared small windows and arranged regularly. Thus, the cages can be used for fabrication of small semiconductor nanoclusters.

Laser ablation (LA) is a well-known method to produce nanoclusters by ablating material from a solid target.[16] LA usually is performed in vacuum, or sometimes in inert gas such as Ar or more reactive gases such as ammonia or nitrogen. Recently a new variation of LA has been reported whereby the target is immersed in a liquid medium, and the laser beam is focused through the liquid onto the target surface.[17] LA technique has been used to produce nanoclusters of semiconductors[18] and metals.[19]

Nanoclusters of II-V semiconductors have attracted not much attention so far. To the best of our knowledge there are several works on $Cd_3P_2$ nanoclusters fabricated by wet chemistry methods[20] and by thermolysis[21] and alcoholysis of organometallic species.[22] As well, in our recent work[23] we have reported the fabrication and study of the optical properties of the nanoclusters of another II-V semiconductor ($ZnP_2$) incorporated into zeolite Na-X matrix. The present paper is the first study of the nanoclusters of another II-V semiconductor: cadmium tetraphosphide ($CdP_4$). Wet chemistry methods seem to us not to be suitable for production of small II-V nanoclusters due to their high reactivity in water. It is hard to expect their high stability in glass melt as well. Thus, incorporation into zeolite cages and production by pulsed laser ablation seem to us to be ones of the most suitable methods for II-V semiconductor nanoclusters fabrication.

Quantum confinement of charge carriers in nanoclusters leads to new effects in their optical properties. Those are the blue shift of exciton spectral lines originating from the increase of the kinetic energy of charge carriers and the increase of the oscillator strength per unit volume.[24] These effects are quite remarkable when the radius of the nanoparticle is comparable with Bohr radius of exciton in bulk crystal. Incorporation into zeolite pores and laser ablation are quite promising methods for fabrication of small nanoclusters were the quantum confinement plays an important role.

Bulk $CdP_4$ crystal is the direct-gap semiconductor[25] with energy gap 0.908 eV. The lattice symmetry is characterized by the space symmetry group $C_{2h}^5$ (monoclinic syngony). Since the bulk crystal has rather narrow energy gap, the blue shifted spectral lines of $CdP_4$ nanoclusters are expected to be in the visible or near IR spectral region.

## 2. Technology of fabrication of $CdP_4$ nanoclusters. Experimental procedures

The high purity $CdP_4$ bulk crystals and synthetic zeolite of Na-X type were used for fabrication of $CdP_4$ nanoclusters. The framework of zeolite Na-X consists of sodalite cages and supercages with the inner diameters of 8 and 13 Å, respectively. $CdP_4$ nanoclusters are too large to be incorporated into small sodalite cage due to existence of many Na cations. Therefore, it is naturally to assume that only the supercages can be the hosts for the nanoclusters. Zeolite and $CdP_4$ crystals were dehydrated in quartz sealed ampoule in vacuum about $2\times10^{-5}$ mm Hg for 1 h at 400°C. We used 100 mm length ampoule for space separation of semiconductor source and zeolite. The fabrication of samples was carried out in two stages. At the first stage (see Figure1) $CdP_4$ was incorporated into the zeolite matrix through the vapour phase at 676°C in source region and 668°C in zeolite region for 100 h. At the second stage, the inverted temperature gradient was applied: 656°C in source region and 665°C in zeolite region. The duration of the second stage was 40 h. The cooling of ampoule were carried out gradually with mentioned above inverted temperature gradient. The inverted temperature gradient was aimed to clean the surface of zeolite crystals from $CdP_4$ film and large particles that, probably, can appear on the zeolite surface. The stability of structure of lattice of zeolite single crystals was controlled by XRD method. The control showed that the zeolite lattice structure was stable at temperatures mentioned above, i.e. clusters were incorporated into the single crystal zeolite matrix. At the end of incorporation process no free cadmium and phosphorus were observed in the ampoule.

The optical spectra were measured from the samples placed both in vacuum in quartz ampoule and in air. A tungsten-halogen incandescent lamp was used as a light source for the diffuse reflection measurements. An $Ar^+$ laser with wavelength 514.5 nm was used in luminescence and Raman experiments. The absorption spectra of the nanoclusters were obtained from the diffuse reflection spectra by conversion with Kubelka-Munk function $K(\hbar\omega) = [1 - R(\hbar\omega)]^2 / 2R(\hbar\omega)$, where $R(\hbar\omega)$ is the diffuse reflectance normalised by unity at the region of no absorption.

For ablation the pulsed Cu laser ($\lambda$=578.2 nm) was used. The pulse intensity of the focused laser beam was about 1.5 MW/cm$^2$, pulse duration was of 20 ns at a repetition rate of 10 kHz. The beam was focused on the surface of target to a spot-size diameter of approximately 0.5 mm. During the ablation the target (bulk CdP$_4$ crystal of area 9 mm$^2$) was dipped into the liquid nitrogen. The produced by ablation nanoclusters of CdP$_4$ were deposited on quartz plate that was positioned on the distance of about 0.2 mm from the surface of the target crystal. The run time of ablation was 20 min. The photoluminescence spectrum was measured from the deposited film excited by cw Ar$^+$ laser ($\lambda$=514.5 nm).

## 3. Structure and optical properties of CdP$_4$ nanoclusters

Diffuse reflection (DR) and photoluminescence (PL) spectra of the CdP$_4$ nanoclusters incorporated into the pores of zeolite Na-X were measured at room (293 K), liquid nitrogen (77 K) and liquid helium (4.2 K) temperatures. Then, the DR spectrum was converted to absorption using the Kubelka-Munk method described above. Within the accuracy of determination of bands spectral positions we did not observe noticeable change of both absorption and PL spectra with temperature. Optical spectra of the clusters were the same both in vacuum in quartz ampoule and exposed to the normal atmosphere condition. This is an evidence of the stability of CdP$_4$ clusters in the pores of zeolite placed in air. The absorption spectrum obtained by Kubelka-Munk method is presented in Figure 2. The spectrum demonstrates two-band structure. The spectral positions of the respective bands signed as B$_1$ and B$_2$ are presented in Table 1. Both bands are blue shifted (see Table 1), where the shift values are calculated using the energy gap of the bulk CdP$_4$ crystal (0.908 eV). Note that the blue shift of the high-energy absorption band for CdP$_4$ clusters in Na-X zeolite is considerably larger than respective one for ZnP$_2$ clusters in the same zeolite [23]: 1.353 eV for CdP$_4$ and 0.808 eV for ZnP$_2$. The observed blue shift allows us to attribute these bands to the absorption into the first electronic excited state of CdP$_4$ nanoclusters incorporated into supercages of the zeolite. The photoluminescence spectrum of $CdP_4$ clusters in zeolite (Figure 3(*a*)) shows the same structure as the absorption one, i.e. PL spectrum consists of the corresponding two $B'_1$ and $B'_2$ bands. Their spectral positions are presented in Table 1. PL bands of nanoclusters are blue shifted as well. The observed blue shift of the absorption and luminescence bands is the result of the quantum confinement of electrons and holes in $CdP_4$ nanoclusters.

Besides the clusters in zeolite pores, we fabricated the $CdP_4$ nanoclusters by pulsed laser ablation technique described above. The obtained PL spectrum of $CdP_4$ clusters is shown in Figure 3(b). The spectrum consists of a single band. One can see that the spectral position of maximum of band, signed as $B''_2$, coincides with the position of $B'_2$ band[1] of luminescence spectrum of $CdP_4$ clusters in zeolite. This fact and the proximity of values of the half-widths of $B''_2$ and $B'_2$ bands (0.183 eV and 0.178 eV correspondingly) allows us to assume the same origin of these PL bands.

But, what kind of $CdP_4$ clusters cause an appearance of the bands in optical spectra. It is often observed that nanoclusters with certain number of atoms are characterized by increased stability (ultrastable nanoclusters) and are more abundant in the sample. This effect is well known for the nanoclusters of different types, e.g. for C,[26] Ar,[27] Na,[27] and for nanoclusters of II-VI semiconductors.[28] Our calculations[23] have shown that such stable nanoclusters exist for $ZnP_2$ that is semiconductor of the II-V type as well as $CdP_4$. Those are $(ZnP_2)_6$ and $(ZnP_2)_8$ with binding energies 1.72 eV and 2.14 eV per atom respectively. Thus, it would be naturally to assume that similarly to $ZnP_2$ the respective stable $(CdP_4)_n$ nanoclusters exists for $CdP_4$ as well. We performed the calculations aimed to find such stable $(CdP_4)_n$ clusters. We considered the stoichiometric clusters only, since no free cadmium and phosphorus were observed in the ampoule as it was mentioned above. Initially, we performed the geometry optimization of the structure of clusters by molecular mechanics MM+ method. Then, we performed the *ab initio* calculation (by STO-3G basis set) of the ground state energy of the clusters with optimized structure. The results are presented in Figure 4. One can see that $(CdP_4)_4$ cluster is the most stable, and clusters with n=1, 3, 5 have the close values of the binding energy, namely $(CdP_4)_4$ has the binding energy of 4.37 eV per atom, $(CdP_4)_1$ – 3.84 eV, $(CdP_4)_3$ – 3.78 eV, $(CdP_4)_5$ – 3.77 eV. The clusters with $n = 2$, 6 and 7 have considerably lower binding energy. Thus, the $(CdP_4)_n$ clusters are bound considerably stronger than the respective $(ZnP_2)_n$ ones. It is naturally to assume that some other stable $(CdP_4)_n$ clusters with $n > 7$ exist as well. However, our estimation of the diameter of largest $(CdP_4)_n$ cluster that might be placed in zeolite Na-X supercage is 9.28 Å. Our estimations take into account the van der Waals radii of Cd (1.940 Å) and P (1.784 Å). Therefore, the clusters with $n \geq 6$ can not be incorporated into supercages of Na-X zeolite. Here and everywhere in the article, maximum diameter of cluster means the distance between the centers of outermost atoms of cluster. Since the $(CdP_4)_4$ cluster is the most stable, it is quite reasonable to assume that this cluster is the most abundant. Correspondingly, $(CdP_4)_4$ cluster would be formed in prevalent quantities at the ablation. An effect of the prevalence of the most stable nanoclusters in mass spectra is well known for II-VI semiconductors.[28] Meanwhile, laser ablation is the method for production of analyzed particles in mass spectrometry. It is naturally that other $CdP_4$ nanoclusters, besides $(CdP_4)_4$, would be formed at the ablation as well, but, probably, their quantity is quite small compared to quantity of $(CdP_4)_4$ ones. Therefore, these less abundant clusters would not give the considerable contribution to PL spectrum. Thus, it is naturally to assume that $B_2''$ band of the PL spectrum of the clusters fabricated at the ablation originates from the $(CdP_4)_4$ cluster. As we have noted above the $B_2'$ band of the PL spectrum of clusters incorporated into zeolite has the same origin as $B_2''$ band. Therefore, the low-energy luminescence $B_2'$ band and the respective absorption $B_2$ one originate from the most stable $(CdP_4)_4$ cluster. And what is the origin of the high-energy absorption $B_1$ and the respective luminescence $B_1'$ bands of clusters in zeolite? Since these bands are characterized by higher energies, they would originate from the clusters smaller than $(CdP_4)_4$ one. Those can be $(CdP_4)_1$ and $(CdP_4)_3$, which are characterized by close values of binding energy.

---

[1] Taking into account rather half-widths of $B'_2$ and $B''_2$ bands, the small (0.01 eV) discrepancy in their maxima positions is negligible.

To clarify the problem of the origin of $B_1$ and $B_1'$ bands we performed Raman study of the clusters in zeolite. The measured Raman spectrum is shown in Figure 6. To answer the question what kind clusters are formed in zeolite pores we performed semi-empirical calculation (by AM1 method) of the vibrational spectrum of the CdP$_4$ clusters that may fit into the zeolite supercage. Calculations showed that the vibrational spectra of only (CdP$_4$)$_4$ and (CdP$_4$)$_3$ clusters agree in satisfactory way with experimental Raman spectrum. Therefore, the Raman spectrum shows that (CdP$_4$)$_4$ and (CdP$_4$)$_3$ clusters are formed in zeolite pores. Therefore, we can conclude that $B_1$ and $B_1'$ bands originate from the absorption and emission transitions respectively in (CdP$_4$)$_3$ cluster. The presence of $B_1'$ band originating from (CdP$_4$)$_3$ cluster in PL spectrum of clusters in zeolite is due to difficulty of penetration of CdP$_4$ substance into most inner areas of zeolite crystals. Therefore, in these areas the smaller clusters with lower binding energy would be formed. Thus, $B_1$ and $B_1'$, $B_2$ and $B_2'$ ($B_2''$) bands can be attributed to originate from mentioned above stable (CdP$_4$)$_3$ and (CdP$_4$)$_4$ nanoclusters. The calculated structure of these clusters is presented in Figure 5. One can see from the figure that due to difference of the covalent radii of Cd and P atoms the (CdP$_4$)$_3$ and (CdP$_4$)$_4$ clusters have belt-like structure.

One can see from the Table 1 that the luminescence bands have the Stokes shift from the absorption ones. The values of this shift are large enough: 0.684 eV for $B_1'$ band and 0.505 eV for $B_2'$ one. These values are considerably larger than ones for ZnP$_2$ nanoclusters in the same Na-X zeolite (0.078-0.135 eV).[23] Stokes shift is well known both in the molecular spectroscopy and in the spectroscopy of nanoclusters. It is known that this kind of Stokes shift (so-called Frank-Condon shift) is due to vibrational relaxation of the excited molecule or nanoparticle to the ground state. The theory of Frank-Condon shift in nanoclusters was developed by Franceschetti and Pantelides[29] where the first-principle calculations of excited-state relaxations in nanoclusters were performed. As they showed, for small nanoclusters the Stokes shift is the Frank-Condon one, which is the result of the vibrational relaxation of the nanoparticle in the excited electronic state. The considerable values of Stokes shift in (CdP$_4$)$_n$ clusters mean the substantial role of vibrational relaxation in excited nanoparticles.

## 4. Conclusions

In conclusion, subnanometer $(CdP_4)_n$ clusters have been prepared both by incorporation into zeolite Na-X pores from the vapor phase and by laser ablation in liquid nitrogen media. Absorption and photoluminescence spectra of clusters in zeolite and PL spectrum of clusters obtained by LA method demonstrate quite large blue shift. The semi-empirical search of the stable structures and the *ab initio* calculations of the binding energy of $(CdP_4)_n$ clusters have been performed up to $n = 7$. The results of spectroscopic study and calculations have allowed us to assume the creation of $(CdP_4)_n$ clusters with $n = 3$ and 4 in zeolite pores and clusters with $n = 4$ in LA experiments. The Raman spectrum of the clusters in zeolite has proved the creation of $(CdP_4)_3$ and $(CdP_4)_4$ clusters in zeolite pores.

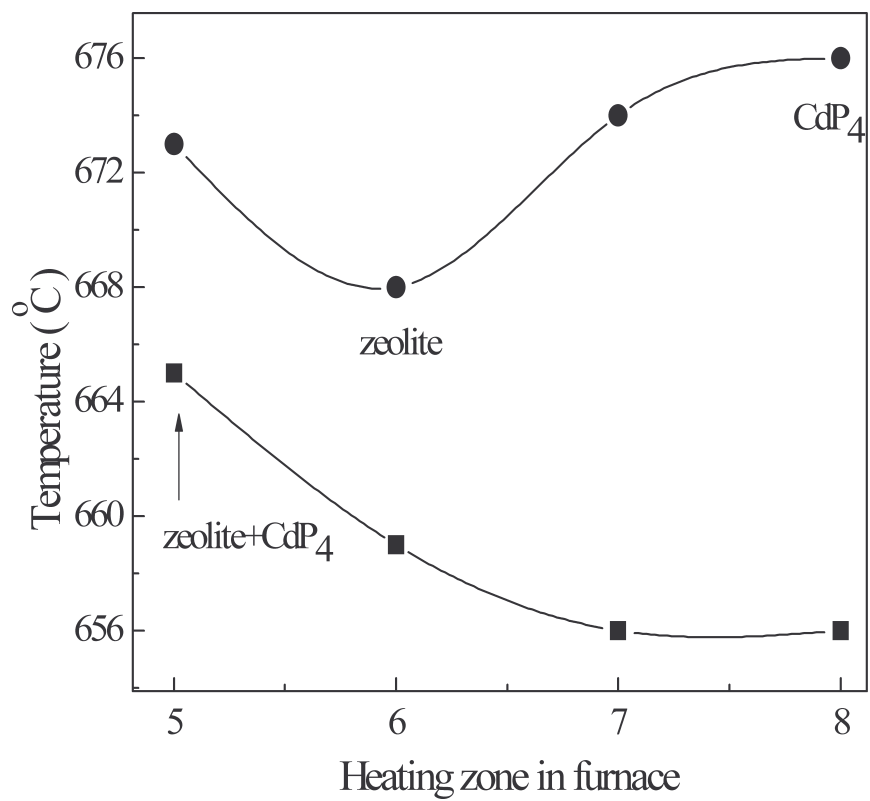

**Figure 1.** The temperature gradients used at first stage (connected filled circles) and at second stage (connected filled squares) of the fabrication of $CdP_4$ clusters in zeolite Na-X.

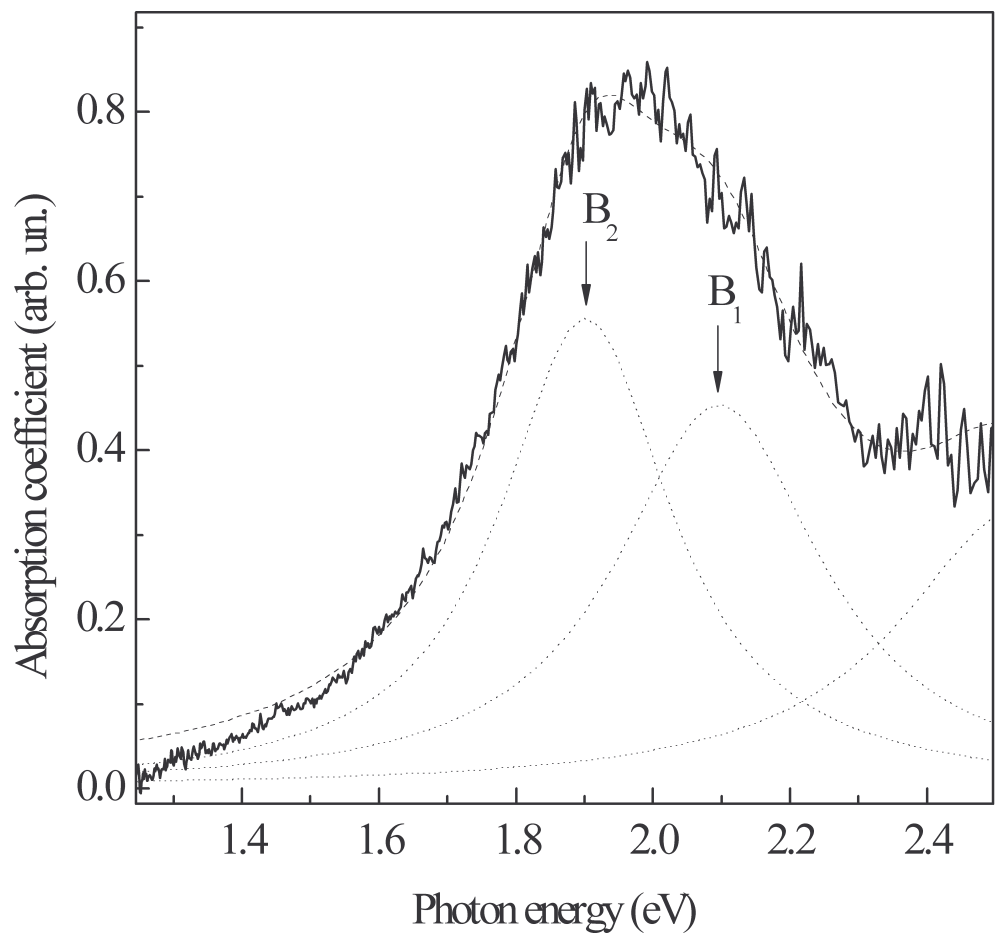

**Figure 2.** The absorption spectrum of $CdP_4$ clusters in zeolite Na-X at the temperature of 77 K.

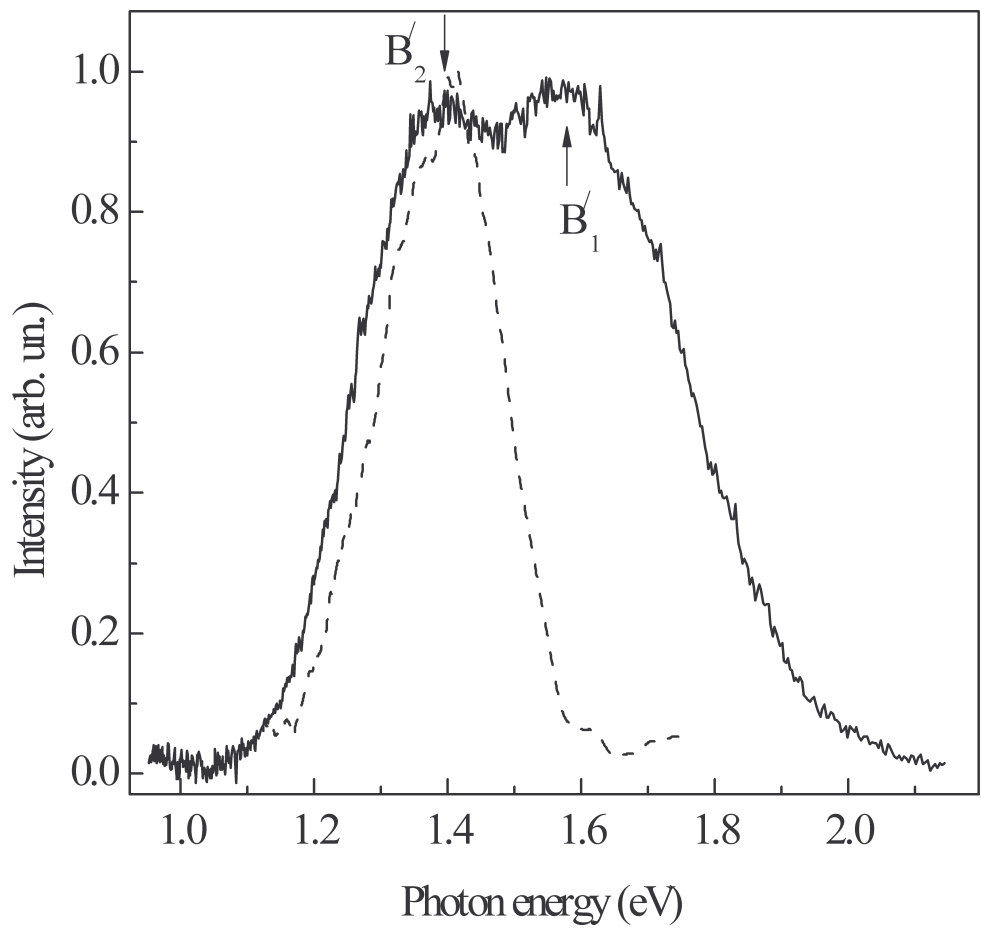

**Figure 3.** The photoluminescence spectrum of $CdP_4$ clusters in zeolite Na-X (solid curve) and the PL spectrum of $CdP_4$ clusters fabricated by laser ablation (dashed curve) at the temperature of 77 K.

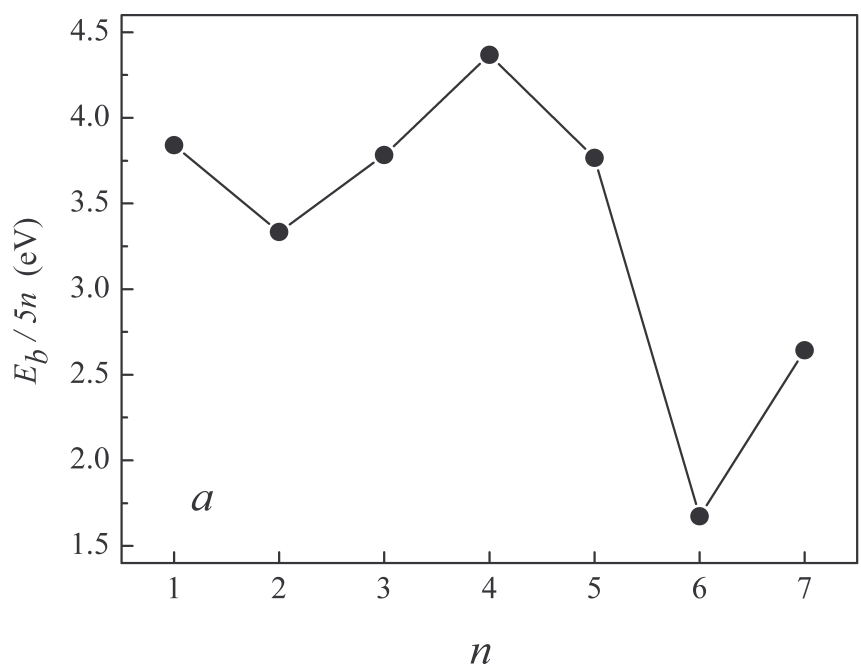

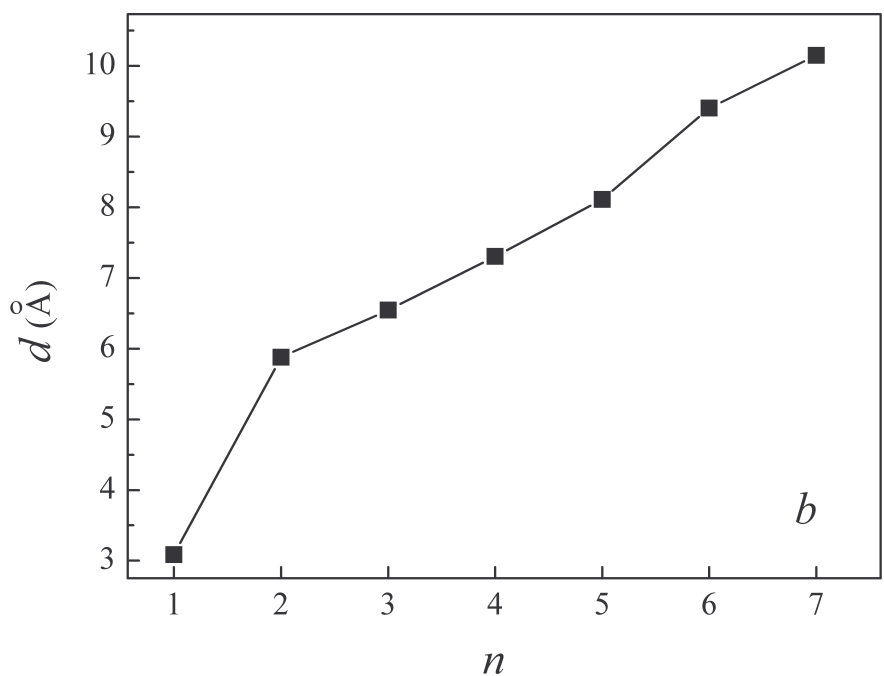

**Figure 4.** The results of the *ab initio* calculations of (CdP$_4$)$_n$ cluster binding energy per atom (*a*) and maximum diameter of cluster (*b*) versus *n*.

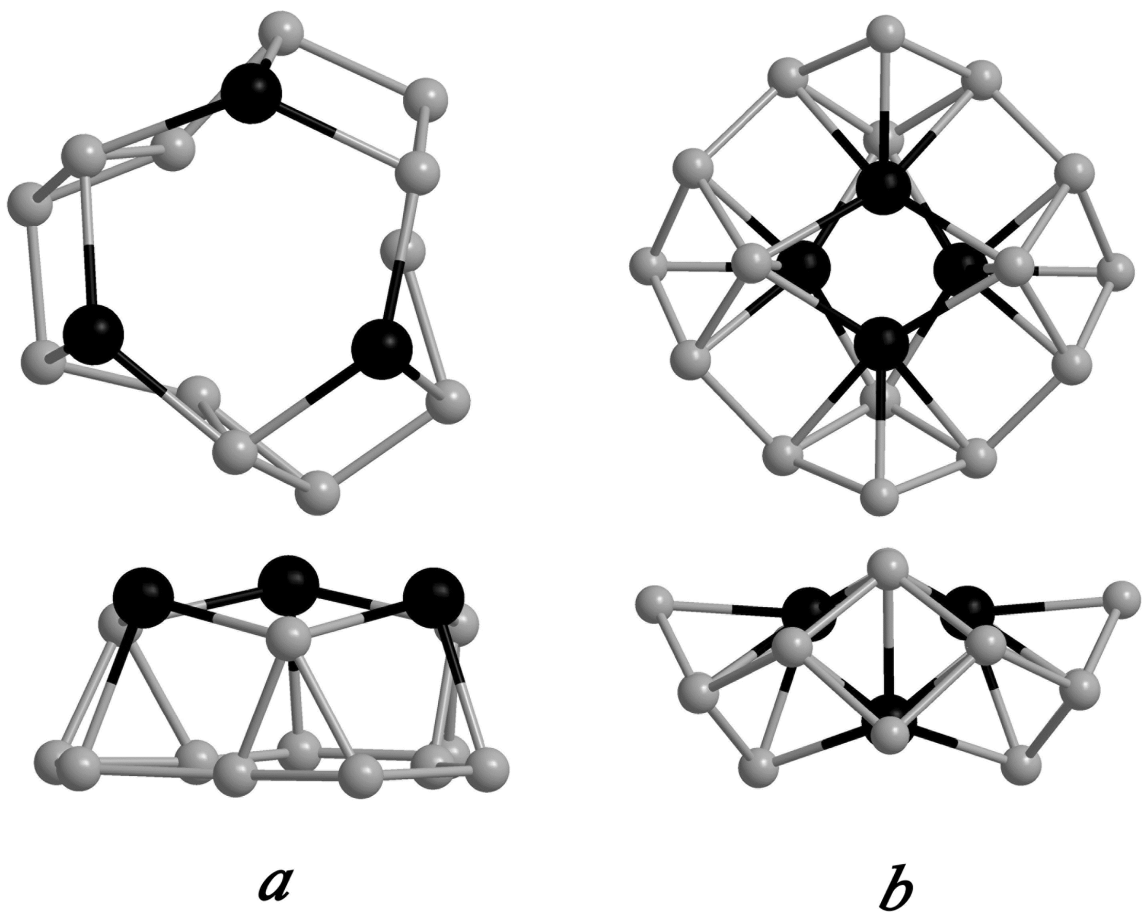

**Figure 5.** Calculated structure of $(CdP_4)_n$ clusters: (*a*) – structure of the $(CdP_4)_3$ cluster, and (*b*) – structure of the $(CdP_4)_4$ one, where Cd atoms – black balls, P atoms – grey balls.

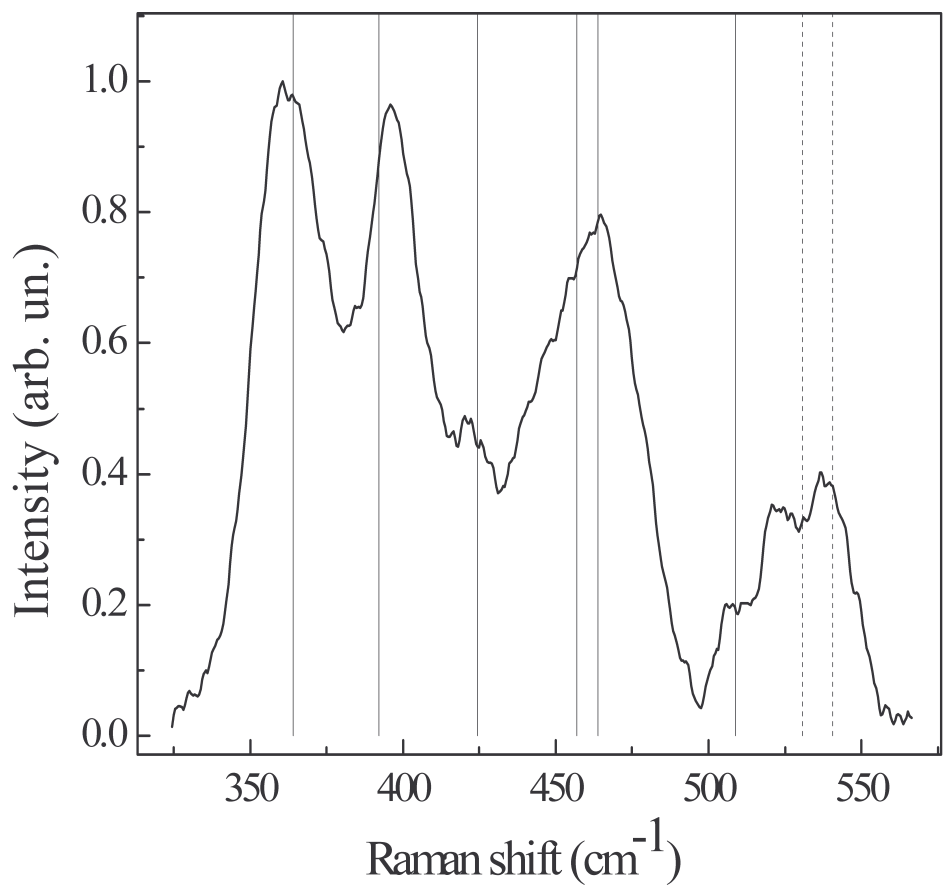

**Figure 6.** The Raman spectrum of CdP$_4$ clusters in zeolite Na-X at the temperature of 293 K. The solid lines mark the calculated vibrational frequencies of (CdP$_4$)$_4$ cluster and the dashed lines mark the respective frequencies of (CdP$_4$)$_3$ one.

**Table 1.** Spectral characteristics of CdP$_4$ clusters in zeolite Na-X and clusters produced by laser ablation.

| Spectral position (eV) | | | Blue shift of absorption band (eV) | Stokes shift (eV) | |
|---|---|---|---|---|---|
| Absorption | PL | | | Zeolite Na-X | Ablation |
| | Zeolite Na-X | Ablation | | | |
| 2.261 (B$_1$) | 1.577 (B$_1'$) | | 1.353 (B$_1$) | 0.684 (B$_1$-B$_1'$) | |
| 1.902 (B$_2$) | 1.397 (B$_2'$) | 1.407 (B$_2''$) | 0.994 (B$_2$) | 0.505 (B$_2$-B$_2'$) | 0.495 (B$_2$-B$_2''$) |